In loving memory of a wonderful grandma whose courageous battle with cancer motivated me to write this paper

# Efficient Meningioma Tumor Segmentation Using Ensemble Learning

Mohammad Mahdi Danesh Pajouh [1][0009-0009-7039-9233]     Sara Saeedi

University of Calgary

## Abstract

Meningiomas represent the most prevalent form of primary brain tumors, comprising nearly one-third of all diagnosed cases. Accurate delineation of these tumors from MRI scans is crucial for guiding treatment strategies, yet remains a challenging and time-consuming task in clinical practice. Recent developments in deep learning have accelerated progress in automated tumor segmentation; however, many advanced techniques are hindered by heavy computational demands and long training schedules, making them less accessible for researchers and clinicians working with limited hardware.

In this work, we propose a novel ensemble-based segmentation approach that combines three distinct architectures: (1) a baseline SegResNet model, (2) an attention-augmented SegResNet with concatenative skip connections, and (3) a dual-decoder U-Net enhanced with attention-gated skip connections (DDUNet). The ensemble aims to leverage architectural diversity to improve robustness and accuracy while significantly reducing training demands. Each baseline model was trained for only 20 epochs and Evaluated on the BraTS-MEN 2025 dataset. The proposed ensemble model achieved competitive performance, with average Lesion-Wise Dice scores of 77.30%, 76.37% and 73.9% on test dataset for Enhancing Tumor (ET), Tumor Core (TC) and Whole Tumor (WT) respectively. These results highlight the effectiveness of ensemble learning for brain tumor segmentation, even under limited hardware constraints. Our proposed method provides a practical and accessible tool for aiding the diagnosis of meningioma, with potential impact in both clinical and research settings.

---

[1] Corresponding Author: Mohammadmahdi.danesh@ucalgary.ca

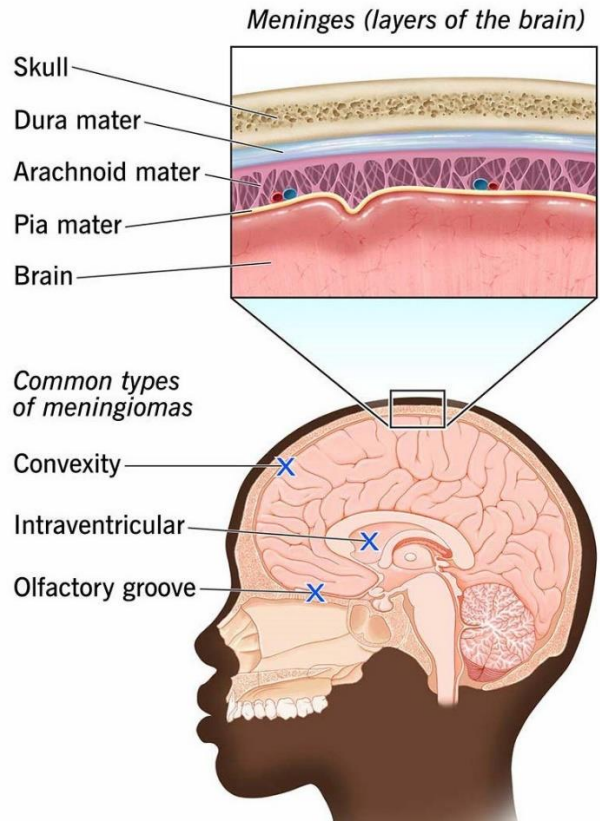

*Figure 1 Visualization of Meninges layers and its common types. Image is from Cleveland Clinic [1]*

## 1. Introduction

### 1.1 Clinical Introduction

A meningioma is a tumor that forms in the meninges, which are three layers of tissue that cover and protect the brain and spinal cord. Meningiomas originate specifically from arachnoid cells, which are found in the thin, spiderweb-like membrane surrounding the brain and spinal cord. This membrane is one of the three layers that make up the meninges shown in figure 1.

Most meningiomas are not cancerous (benign), although they can sometimes be malignant (cancerous). In general, if a tumor is cancerous, it means it is aggressive, can invade other tissues, and potentially spread to other parts of the body. A benign tumor, on the other hand, does not spread.

Meningiomas are most often found near the top and outer curve of the brain. They may also form at the base of the skull. Spinal meningiomas are rare. Meningiomas tend to grow slowly and inward. Often, they have grown quite large before being diagnosed. Even benign meningiomas can become life-threatening if they compress or affect nearby areas of the brain. There are three types of meningioma based on grade:

- **Grade I (typical):** A benign meningioma that grows slowly. These tumors represent approximately 80% of cases.
- **Grade II (atypical):** A noncancerous meningioma that grows more quickly and can be more resistant to treatment. These account for about 17% of cases.
- **Grade III (anaplastic):** A malignant (cancerous) meningioma that grows and spreads quickly. These represent approximately 1.7% of cases [1].

Accurate and early segmentation of meningiomas from MRI is essential for diagnosis, treatment planning (e.g., surgical resection or radiation therapy), and longitudinal monitoring. However, their variable location and morphology make automated segmentation a challenging task.

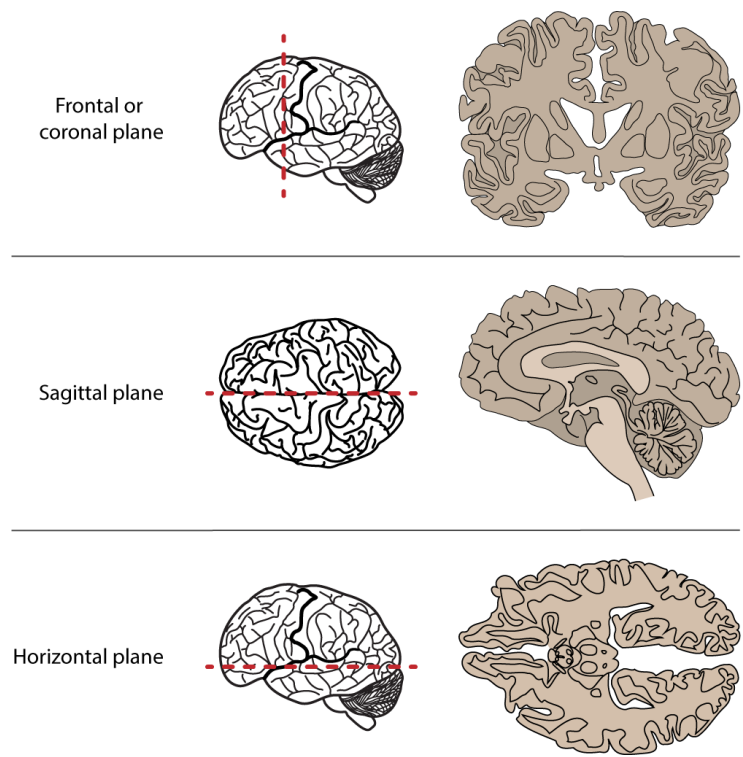

*Figure 2 Representation of the three principal orientations commonly employed in MRI. The upper row illustrates the coronal (frontal) view, which sections the brain from side to side, separating anterior from posterior regions. The middle row displays the sagittal perspective, slicing the brain lengthwise to distinguish the left and right hemispheres. The lower row shows the axial (horizontal) orientation, which produces top-to-bottom slices spanning superior to inferior aspects. Taken together, these planes provide complementary insights into brain anatomy and are widely utilized in both diagnostic practice and neuroscientific research. Image adapted from Foundations of Neuroscience by Casey Henley (2021), licensed under CC BY-NC-SA 4.0.*

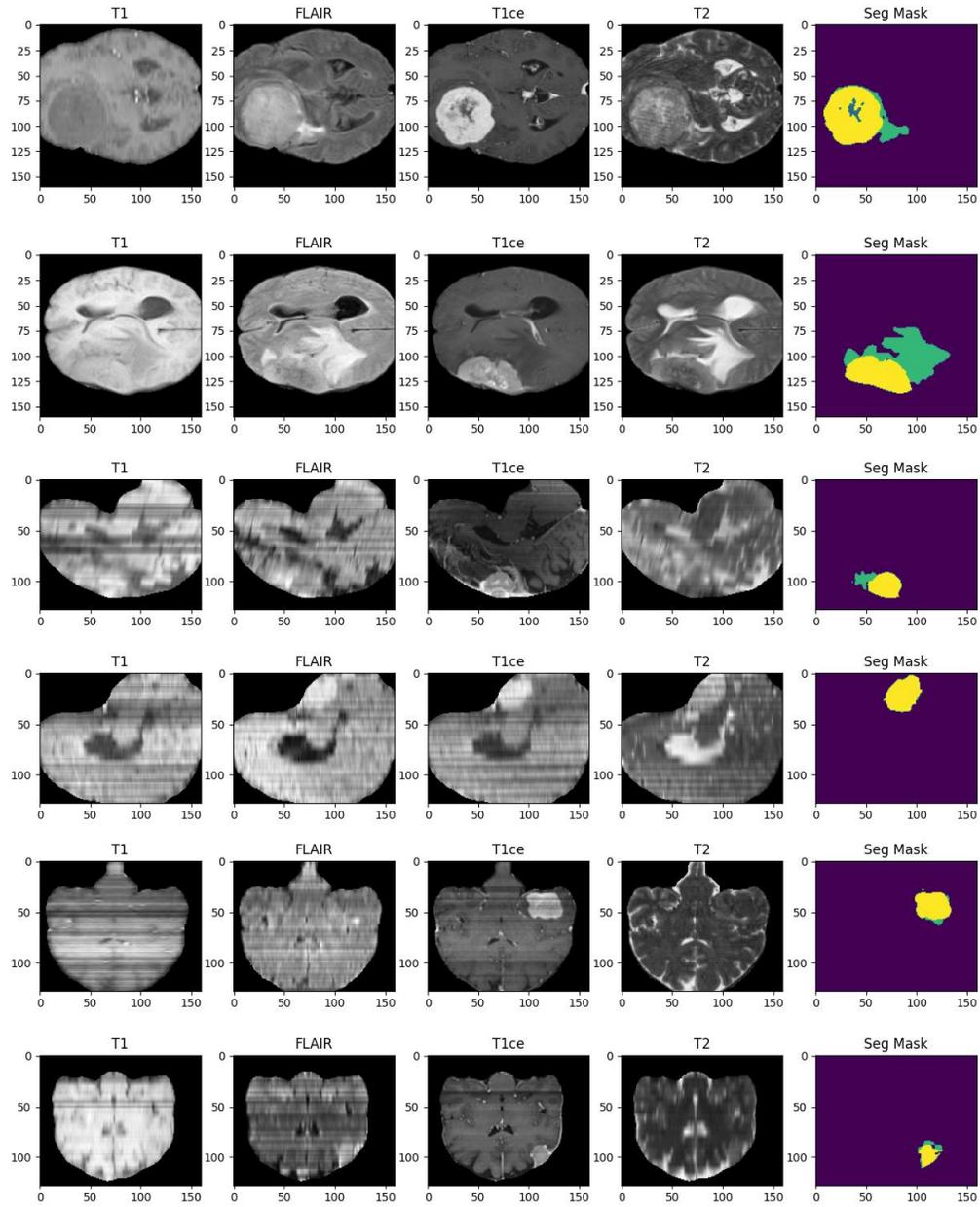

*Figure 3 Illustration of multi-modal MRI scans from the BraTS-MEN 2025 dataset. For each patient, representative horizontal, sagittal, and axial views are shown across four imaging sequences: T1-weighted (T1), Fluid-Attenuated Inversion Recovery (FLAIR), contrast-enhanced T1-weighted (T1ce), and T2-weighted (T2). The final column illustrates the corresponding ground-truth segmentation. Together, these modalities highlight complementary structural and contrast patterns that are essential for reliable identification of tumor regions.*

## 1.2 Magnetic Resonance Imaging

Magnetic Resonance Imaging (MRI) is a powerful, non-invasive diagnostic tool used extensively in clinical and research settings to obtain high-resolution anatomical images. By detecting variations in the alignment of hydrogen nuclei in response to magnetic fields and radiofrequency pulses, MRI enables visualization of internal tissues, particularly those with high water content [2]. In the context of brain tumors, clinicians typically rely on a combination of MRI sequences, each tailored to highlight different tissue properties.

The most frequently used MRI modalities in brain tumor evaluation include:

- **T1-weighted (T1):** Provides clear anatomical structure and is useful for visualizing normal brain anatomy.
- **T1-weighted post-contrast (T1ce):** Acquired after the administration of a gadolinium-based contrast agent, this sequence enhances visualization of vascularized regions, such as tumor tissue or inflammation.
- **T2-weighted (T2):** Highlights fluid-rich areas, making it ideal for detecting edema and cystic regions.
- **FLAIR (Fluid-Attenuated Inversion Recovery):** Suppresses the signal from cerebrospinal fluid, thereby enhancing the visibility of lesions adjacent to ventricles and sulci.

These modalities work synergistically to offer a multi-faceted view of tumor characteristics, aiding in diagnosis, treatment planning, and monitoring of progression.

MRI data is acquired as volumetric 3D scans, consisting of stacked 2D slices. For interpretation and analysis, images are typically viewed in three standard anatomical planes depicted in figure 2:

- **Axial View:** Horizontal slices from top to bottom of the brain.
- **Coronal View:** Vertical slices dividing the brain into anterior (front) and posterior (back) halves.
- **Sagittal View:** Vertical slices that split the brain into left and right sections.

This multi-planar approach provides essential spatial context for accurately assessing tumor size, location, and interaction with surrounding brain structures, which is critical for segmentation tasks.

## 1.3 Dataset

The BraTS-MEN 2025 dataset visualized in figure 3, which is the same as the 2023 version, is a widely recognized benchmark in brain tumor segmentation research. It provides multi-modal MRI scans accompanied by detailed manual annotations that delineate key tumor subregions.

These annotations distinguish tissue characteristics critical for clinical interpretation. Specifically:

- **Label 0** represents background,
- **Label 1** denotes non-enhancing tumor core, including necrotic regions and cystic changes or calcifications,
- **Label 2** corresponds to the surrounding FLAIR hyperintensity, capturing the full extent of abnormal FLAIR signal not part of the core,
- **Label 3** indicates enhancing tumor, reflecting actively growing or vascularized tumor tissue.

Researchers often aggregate these labels into clinically meaningful regions:

- **Whole Tumor (WT):** Union of labels 1, 2, and 3,
- **Tumor Core (TC):** Union of labels 1 and 3,
- **Enhancing Tumor (ET):** Label 3 only.

By offering comprehensive and standardized annotations, the BraTS dataset supports consistent evaluation across segmentation models and facilitates the development of algorithms capable of accurately distinguishing between tumor components [3, 4].

### 1.4 Proposed Model

As a result of thorough experimentation with various architectures and approaches, we found that the best performance was achieved through an ensemble of models. For this task, the ensemble includes the following components:

- **MONAI's SegResNet [5]**, which is essentially a U-Net variant with residual connections,
- A **modified SegResNet** with attention gates in the skip connections and concatenation instead of summation,
- A modified version of **attention-based dual-decoder U-Net (DDUNet)** previously introduced in [6].

All models in the ensemble are evaluated on validation set by BraTS challenge using lesion-wise dice score and hausdorff95.

The rationale for proposing this method is twofold, reflecting both performance-oriented and practical objectives:

**1. Achieving Accurate Segmentation**

The proposed ensemble was achieved through experimenting with different architectural designs, hyperparameters, and training strategies. The aim was to identify a solution that provides robust segmentation performance while keeping computational demands low. Among all tested configurations, the ensemble approach consistently outperformed individual models on the validation set.

**2. Promoting Accessibility and Hardware Efficiency**

A primary motivation behind this work is to develop a model that is accessible to users with limited computational resources. Many high-performing models in the literature require GPUs with large memory and long training times, making them impractical for many researchers and clinicians. In contrast, our ensemble was trained and tested on modest hardware (e.g., a standard home PC), for a few epochs, yet still delivered strong results. This makes it feasible for others to reproduce, fine-tune, or deploy the model without requiring high-end infrastructure.

## 2. Methodology

### 2.1 Overview

To address the task of meningioma segmentation, we adopt an ensemble-based strategy that integrates MONAI's implementation of SegResNet, an attention residual UNet which is different from SegResNet due to addition of attention gate modules in skip connections and using concatenation of encoder features (through skip connections) and decoder features, unlike SegResNet that uses summation. Lastly a modified version of attention dual-decoder UNet (DDUNet) [6] was used. The DDUNet in this work differs from original by incorporating channel-wise attention (squeeze-and-excitation) and adding residual connections. Adding this attention module empirically improved the performance. The attention gate module was used in the last two models, allowing the networks to prioritize tumor-relevant regions while suppressing irrelevant background information. The models were trained using a combination of dice loss and focal loss which helps mitigate class imbalance in the dataset. To further promote generalization, we incorporated 3D dropout and weight decay during optimization. We utilize the BraTS-MEN 2025 dataset with four MRI modalities.

### 2.2 Preprocessing

Each subject in the BraTS-MEN 2025 dataset is provided with four volumetric MRI sequences, together with a segmentation mask annotated across four categories (labels 0–3). Raw volumes are stored in 3D format and were converted into NumPy arrays for efficient processing. The modalities were concatenated along the channel dimension to form a unified input tensor.

Because the original scans are large ($240 \times 240 \times 155$ voxels), we applied center cropping to reduce input size and computational cost while preserving anatomical fidelity. This yielded a final crop of $160 \times 160 \times 128$ voxels, which was used across all experiments. Segmentation masks were transformed into one-hot encodings to represent the four target classes. The training dataset contains 1000 subjects, while the validation set contains 141 subjects.

Intensity normalization was performed on each subject individually using z-score standardization, ensuring consistent intensity distributions across patients. To improve generalization and reduce overfitting chance, several random augmentations are applied to each training sample for the first epochs of DDUNet model such as: random affine scaling and rotation, random Gaussian noise, blurring and scaling intensities. This was done only on DDUNet because it has more parameters compared to the other two models. Adding augmentation to the other two, did not improve performance.

### 2.3 Model Architectures

The ensemble is designed to take advantage of several models' strength. It uses three lightweight models that grow in complexity, simple SegResNet which is quite light, a more complex version of residual UNet with attention and the highest number of parameters belongs to residual DDUNet with channel-wise attention. We will discuss each one in more detail.

**SegResNet**

We employ the MONAI's implementation of SegResNet (without variational autoencoder regularization) as our first model. There are four encoder layers with 1, 2, 2 and 4 residual blocks per each level respectively. The decoder has three levels each containing 1 residual block. The dropout probability is set to 0.2 and the initial number of filters is 16.

**Attention Residual UNet (Modified SegResNet)**

We implemented this model by overriding MONAI's SegResNet class and adding attention gate mechanisms at each skip connection (explained in detail later in this subsection). Unlike SegResNet, we decided to concatenate the features coming from encoder through skip connections and decoder features. This provides the model with more flexibility to choose from the features. Number of blocks in encoder and decoder, dropout rate and initial filters is the same as our SegResNet model.

**Modified DDUNet**

DDUNet is the result of our previous work, in which more than 100 training strategies and architectures were tested for the task of glioma brain tumor segmentation. The model was modified and tailored to the new task of meningioma tumor segmentation. We found that adding channel-wise attention modules improve the performance. In addition, the convolution blocks in DDUNet were replaced by residual blocks. A short summary of the modified DDUNet is presented below:

Encoder

The encoder portion adopts a five-stage hierarchical design similar to U-Net, where each level consists of a residual block containing between two and four 3D convolutions (depending on depth). Every convolution is followed by Group Normalization and a ReLU activation. Group Normalization [7] was selected in place of the more common Batch Normalization [8] because training is performed with a batch size of one, where group-based statistics provide greater stability. Feature channel dimensions expand gradually across layers as follows:

$$\text{Input}=4 \rightarrow 16 \rightarrow 32 \rightarrow 64 \rightarrow 128 \rightarrow 256.$$

Dual Decoders

Instead of a single decoding path, the network uses two separate decoder streams. Each branch performs upsampling and merges its activations with the attention-weighted skip features from the encoder. The merged tensors are then refined with a residual block.

- Decoder 1 is composed of four, four, three, and two convolutional blocks across its levels.
- Decoder 2 includes three, three, two, and two convolutional blocks.

Both branches produce independent segmentation outputs. These predictions are concatenated along the channel dimension and passed through a concluding 1×1×1 convolution to yield the final voxel-level segmentation map. This late fusion step enables the model to exploit complementary feature representations learned by the two decoders.

Normalization and Regularization

All residual units utilize Group Normalization, which is well-suited for very small batch sizes (in our case, one). To encourage robustness, 3D dropout with probability 0.1 is applied after each residual block. Unlike standard dropout, Dropout3D discards entire feature maps rather than individual activations, which helps maintain structural consistency, an advantage when segmenting tumors that may be spatially small. Weight decay regularization with weight 0.001 is also applied.

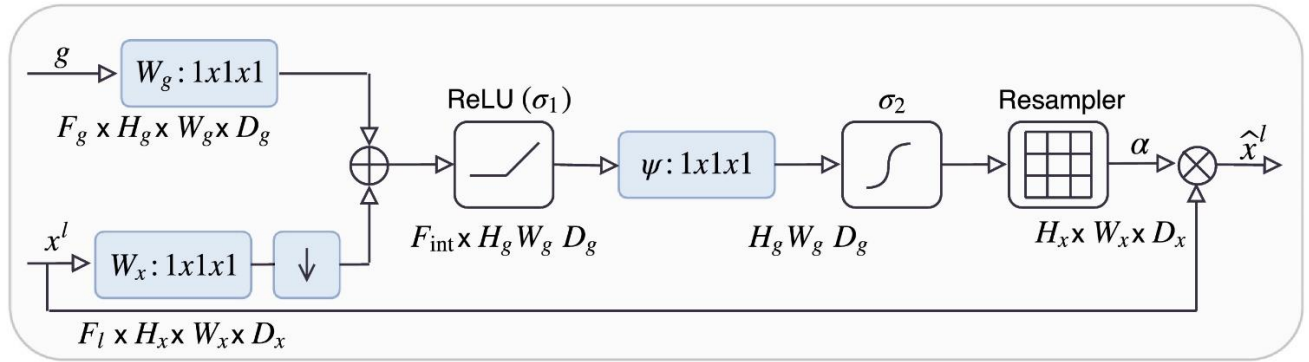

*Figure 4 Illustration of the Attention Gate (AG) as introduced by Oktay et al. in the Attention U-Net [9]. Each gate receives two inputs: a gating signal $g$ from a deeper decoder layer and a skip connection $x^l$ from the encoder. Both inputs are first projected via 1×1×1 convolutions to reduce channel dimensions. The resulting feature maps are combined through element-wise addition, passed through a ReLU, and then processed with another 1×1×1 convolution followed by a sigmoid activation to generate attention coefficients α. These coefficients modulate the encoder features, emphasizing relevant spatial regions.*

*In the original Attention U-Net, a resampling step aligns the spatial dimensions of the gating signal and encoder features. In contrast, our adaptation operates at the same spatial resolution for both inputs, eliminating the need for resampling while retaining fine-grained spatial details.*

To improve the effectiveness of skip connections, attention gate modules are integrated independently into each decoder of the DDUNet as well as into the modified SegResNet models. These gates calculate attention weights for encoder features based on the activations of the corresponding decoder stage. The mechanism follows an additive attention formulation.

In the original Attention U-Net (Oktay et al., 2018) [9] shown in figure 4, the gating signal is obtained from a deeper, lower-resolution decoder layer, which modulates the encoder features prior to fusion. While this helps suppress irrelevant background signals, it may also compromise the preservation of fine spatial details.

Our DDUNet variant employs same-level gating, where the gating signal is derived from decoder features at the same spatial resolution as the encoder skip connections. The encoder features (X) and decoder signal (G) are first projected into a shared feature space through 1×1×1 convolutions. Their combination is passed through a ReLU activation, followed by another 1×1×1 convolution with a sigmoid function to produce the attention map. This map is then applied to reweight the encoder features before they are concatenated with the decoder outputs. By aligning gating and encoder features at the same resolution, the network preserves more local detail, which improves segmentation of small or complex tumor structures.

**Loss Function**

To handle the inherent class imbalance and enhance segmentation accuracy across tumor subregions, we employ a hybrid loss that integrates multi-class Dice Loss with multi-class Focal Loss. Dice Loss encourages high overall overlap with ground truth masks, while Focal Loss emphasizes voxels that are more difficult to classify, effectively guiding the model to focus on challenging regions. The combination ensures robust performance across all tumor classes by balancing global structure alignment with fine-grained voxel-level attention.

Multi-Class Dice Loss

The Dice similarity coefficient is a widely used metric for evaluating segmentation performance, particularly in cases where class imbalance is prevalent. In multi-class segmentation, the Dice score is calculated separately for each class, and the overall loss is derived by averaging the class-wise scores. This approach helps ensure that smaller tumor subregions receive sufficient gradient contribution during training. The Dice loss is expressed as:

$$L_{Dice} = 1 - 2 \times \frac{P \cap T}{P + T}$$

where $P$ is the predicted set and $T$ is the ground truth.

Multi-Class Focal Loss

Focal loss extends the conventional cross-entropy loss by assigning reduced weights to well-classified voxels and amplifying the impact of harder, misclassified examples. This re-weighting mechanism is especially beneficial in medical segmentation tasks where easily classified background voxels can dominate the loss. The focal loss is defined as:

$$L_{Focal} = -\sum_{n=1}^{N} -\alpha(1 - p_t)^{\gamma} \log(p_t)$$

If ground truth is 1, then $p_t = p$, otherwise $p_t = 1 - p$. $\gamma$ is called the focusing parameter and in our setup it is 2. Additionally, $\alpha = 0.25$ is the parameter for class-balancing.

Final Loss

The two loss terms are combined to form the final loss function:

$$L_{total} = \lambda_1 \times L_{Dice} + \lambda_2 \times L_{Focal}$$

In our experiments, we use $\lambda_1 = 0.75, \lambda_2 = 0.25$ as they demonstrated superior performance.

### 2.4 Post-processing

After the segmentation masks are created for validation set by getting the majority vote of the result of each model, the images are padded with label 0 to match their original shape. An affine transform is applied to position the masks according to the challenge guidelines.

### 2.5 Training Procedure

We trained our models using the AdamW optimizer with AMSGrad and learning rate of $5 \times 10^{-5}$. A combination of dice loss and focal loss were used to address the class imbalance issue.

Training was carried out with batch size one, using a GTX1080 GPU with 8GB vram.

## 3. Experiments and Results

On validation set, for each subresion namely enhancing tumor (ET), tumor core (TC) and whole tumor (WT), the average and median results of each baseline model and the ensemble is presented in table 1 and table 2 respectively. Throughout our experiments, the ensemble constantly outperformed single models.

*Table 1 Average results of the three baseline and the ensemble models on validation set*

| Model | Lesion-wise Dice ET (%) | Lesion-wise Dice TC (%) | Lesion-wise Dice WT (%) | Lesion-wise hausdorff95 ET | Lesion-wise hausdorff95 TC | Lesion-wise hausdorff95 WT |
|---|---|---|---|---|---|---|
| SegResNet | 73.9 | 74 | 72.4 | 63.29 | 60.2 | 62.6 |
| Modified SegResNet | 73.3 | 75.7 | 67.8 | 73.3 | 63.5 | 92.4 |
| Modified DDUNet | 62.6 | 63 | 61.5 | 109 | 107.1 | 110.8 |
| Ensemble | 76.7 | 76.2 | 73.8 | 56.5 | 55.9 | 63.1 |

Table 2 Median results of the three baseline and the ensemble models on validation set

| Model | Lesion-wise Dice ET (%) | Lesion-wise Dice TC (%) | Lesion-wise Dice WT (%) | Lesion-wise hausdorff95 ET | Lesion-wise hausdorff95 TC | Lesion-wise hausdorff95 WT |
|---|---|---|---|---|---|---|
| SegResNet | 89.6 | 88.2 | 87.3 | 1.9 | 2 | 2.4 |
| Modified SegResNet | 92.4 | 93.3 | 87.6 | 1.4 | 1.6 | 3.6 |
| Modified DDUNet | 79.4 | 79.7 | 77.7 | 5 | 5 | 5.7 |
| Ensemble | 91.8 | 91 | 88.2 | 1.4 | 1.4 | 2.2 |

On the test set, the lesion-wise dice is reported as 77.30%, 76.37% and 73.92% with standard deviations of 0.29, 0.30 and 0.28 for enhancing tumor (ET), tumor core (TC) and whole tumor (WT) respectively.

Through experimentation, we realized that these models complement each other in various samples. For example DDUNet, would usually overestimate tumor regions leading to more false positives. However, adding it to the ensemble improved performance as SegResNet models were too conservative when used alone.

We also observed that sometimes improving the performance of a model causes the performance of the ensemble to deteriorate, which shows there is no guarantee that improving overall performance of a model would improve its contribution to ensemble per sample.

Additionally, there were 9 samples in validation set that all the tested models predicted wrongly. These samples could correspond to cases in which the model has not been trained on.

Another notable aspect of this work is that each baseline model was trained for only 20 epochs, which further highlights a good choice of architecture could result in competitive results in one to two days.

## 4. Conclusion

Meningioma is a very common primary brain tumor. In recent years, AI has proved to be a great tool to diagnose and detect diseases. In the case of brain tumors however, due to the abnormal morphology of tumors, the development of accurate, efficient, and generalizable segmentation models continues to pose significant challenges, particularly in settings with limited computational resources. In this study, we introduced a novel ensemble-based approach composed of three lightweight models: a baseline SegResNet, an attention-augmented SegResNet with concatenative skip connections, and a modified attention dual-decoder U-Net with added channel-wise attention and residual blocks (DDUNet). Each baseline model used in the ensemble was trained for only 20 epochs.

Our ensemble was evaluated on the BraTS-MEN 2025 dataset and achieved competitive performance on unseen test set, with average Lesion-Wise Dice scores of 77.30%, 76.37% and 73.92% for Enhancing Tumor (ET), Tumor Core (TC) and Whole Tumor (WT) respectively. These results demonstrate that it is possible to achieve great segmentation accuracy without relying on large-scale hardware or extensive training schedules.

Beyond accuracy, the strength of our approach lies in its accessibility and practical applicability. By leveraging architectural diversity and attention mechanisms, our method balances performance and efficiency, making it an ideal solution for deployment in clinical or research settings with limited resources.

**Acknowledgment**

This research was supported by University of Calgary. The authors would like to thank Hossein Danesh Pajouh for his encouragement and providing the computer used in this research.